\documentclass{elsart}
\usepackage{amssymb,amsfonts,graphicx}

\begin{document}
\begin{frontmatter}

\title{Grouping in the stock markets of Japan and Korea}

\author[kaist]{Woo-Sung Jung\corauthref{cor1}}
\ead{wsjung@kaist.ac.kr} \corauth[cor1]{Corresponding author. Fax:
+82-42-869-2510.}
\author[kaist]{Okyu Kwon}
\author[icu]{Taisei Kaizoji}
\author[kaist]{Seungbyung Chae}
\author[kaist]{Hie-Tae Moon}
\address[kaist]{Department of Physics, Korea Advanced Institute of Science and Technology,
Daejeon 305-701, Republic of Korea}
\address[icu]{Division of Social Sciences, International
Christian University, Tokyo 181-8585, Japan}

\begin{abstract}

We investigated the temporally evolving network structures of the
Japanese and Korean stock markets through the minimum spanning trees
composed of listed stocks. We tested the validity of conventional
grouping by industrial categories, and found a common trend of
decrease for Japan and Korea. This phenomenon supports the
increasing external effects on the markets due to the globalization
of both countries. At last the Korean market are grouped with the
MSCI Korea Index, a good reference for foreigners' trading, in the
early 2000s \cite{Jung2006}. In the Japanese market, this tendency
is strengthened more and more by burst of the bubble in 1990's.

\end{abstract}


\end{frontmatter}

\section{Introduction}
The study of financial market has received attention from physicists
\cite{Mantegna1999B,Arthur1997B}. Nowadays, the network theory is
used to investigate complex systems with many interacting agents,
and also many physicists have analyzed the financial market through
the network theory. The financial market, where all listed companies
are correlated with each other, is a typical complex systems.

The minimum spanning tree (MST) is useful to construct the asset
tree, and it provides the study to find the market's characteristics
simply
\cite{Mantegna1999,Micciche2003,Onnela2004,Coronnello2005,Bonanno2003,Mizuno2006}.
In the MST of $N$ nodes, each node represents a company, and $N-1$
links with the most important correlations are selected. The MST is
a loop-less network, and all of nodes have at least one link. This
tree is obtained from a fully connected network of the correlation
matrix. Also, the grouping of companies can be identified and
extended to portfolio optimization. The companies of the US market
are clearly grouped with the industry category or business sector
\cite{Onnela2003}.

Several papers that the characteristics of the mature market cannot
be simply extended to emerging markets for every case in the recent
days \cite{Matia2004,Yan2005}. The Korean market is an emerging
market and shows many particular properties including the grouping
method \cite{Jung2006,Koreanmarket1,Koreanmarket2,Koreanmarket3}.
Also, the Japanese market is an attractive market for
econophysicists to study
\cite{Kaizoji2000,Kaizoji2004,Kaizoji2004B}, and there are high
correlations between two East Asian markets, the Japanese and the
Korean market.

Korea and Japan have developed close interdependence of their
economic systems for a long time, which leads to many common
features of their economic systems including financial markets. Two
countries experienced very high rate of economic growth, so Japan is
ranked as the world's second largest economy, and the economy of
Korea as the tenth now. Their economies were commonly driven by the
government-directed investment model and the protective trade policy
at first. The success stories of two countries made East Asian
economic model popular, and attracted much attention until 1990s.
However, the long-term depression happened in Japan in early 1990s
\cite{Kaizoji2004}, and the severe recession happened in Korea due
to the Asian Financial Crisis in late 1990s changed the economic
environment very much \cite{Climent2003}. The integration into the
global economy has much developed in two countries in the recent
days. Such violent changes make the studies of the East Asian
financial markets very interesting, and we investigate
characteristics of the Japanese and Korean stock market with the
history of the markets.

\section{Data analysis}
\begin{figure}
\includegraphics[angle=0,width=1.0\textwidth]{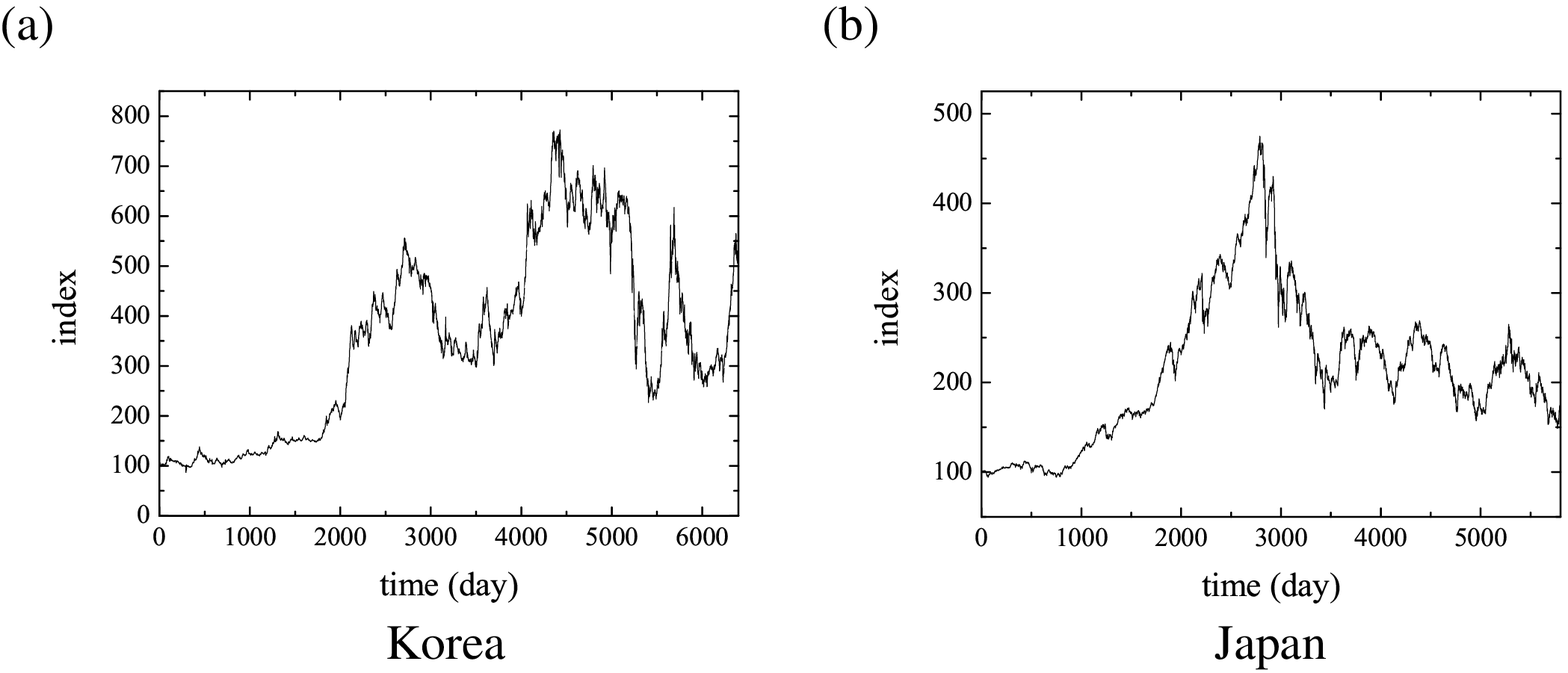}
\caption[0]{Index of selected companies of (a) the Korean stock
market and (b) the Japanese market from 1980 to 2001. }
\label{index}
\end{figure}

We investigated grouping in the Japanese and Korean stock markets.
There are several stock exchange markets in Japan and Korea, and we
selected the Tokyo Stock Exchange (TSE) and the Korea Stock Exchange
(KSE), the largest markets in Japan and Korea, respectively. In
2004, 2306 companies were listed on the TSE, which in its present
form was founded in 1949. The KSE opened in 1956, and 695 companies
are listed in 2005. We used the daily closure stock price from 1980
to 2001, in other words, the time interval ($\Delta t$) used was 1
day. A total of 228 companies in the Korean market and 624 companies
in the Japanese market were selected for our analysis. All of these
companies remained in the markets over this period 21 years. Fig.
\ref{index} shows the index for the companies selected. The
representative TSE and KSE indexes, NIKKEI and KOSPI, are indexes of
the value-weighted average of stock prices. However, the indexes in
Fig. \ref{index} are price-equally-weighted indexes, such as that
used for the Dow Jones Industrial Average (DJIA).

The cross-correlation coefficient between stock $i$ and $j$ is
defined as:

\begin{equation}
\rho_{ij} =\frac{(<S_{i}S_{j}>-<S_{i}><S_{j}>)}
{\sqrt{(<S_{i}^{2}>-<S_{i}>^{2})(<S_{j}^{2}>-<S_{j}>^{2}) }},
\end{equation}

\noindent where $S_i$ is the logarithmic return of a given stock
$i$. The correlation coefficient $\rho_{ij}$ has values from $-1$ to
$1$. The logarithmic return is written as $S_i(t)=\ln Y_i(t+\Delta
t) - \ln Y_i (t)$, where $Y_i(t)$ is the price of a given company
$i$.

We construct the minimum spanning trees (MSTs) with time windows of
width $T$ corresponding to daily data for \emph{3 years}, with
$\delta t$ of approximately 1~month. The MST is a simple graph with
the most important connection selected \cite{Mantegna1999}. The MST
of $N$ nodes has $N-1$ links, and every node has at least one link.
Each node of the network corresponds to a company, and each link has
a weight $w_{ij}(=w_{ji})$, which is simply the value of the
cross-correlation coefficient, $w_{ij}=\rho_{ij}$.

\begin{figure}
\includegraphics[angle=0,width=1.0\textwidth]{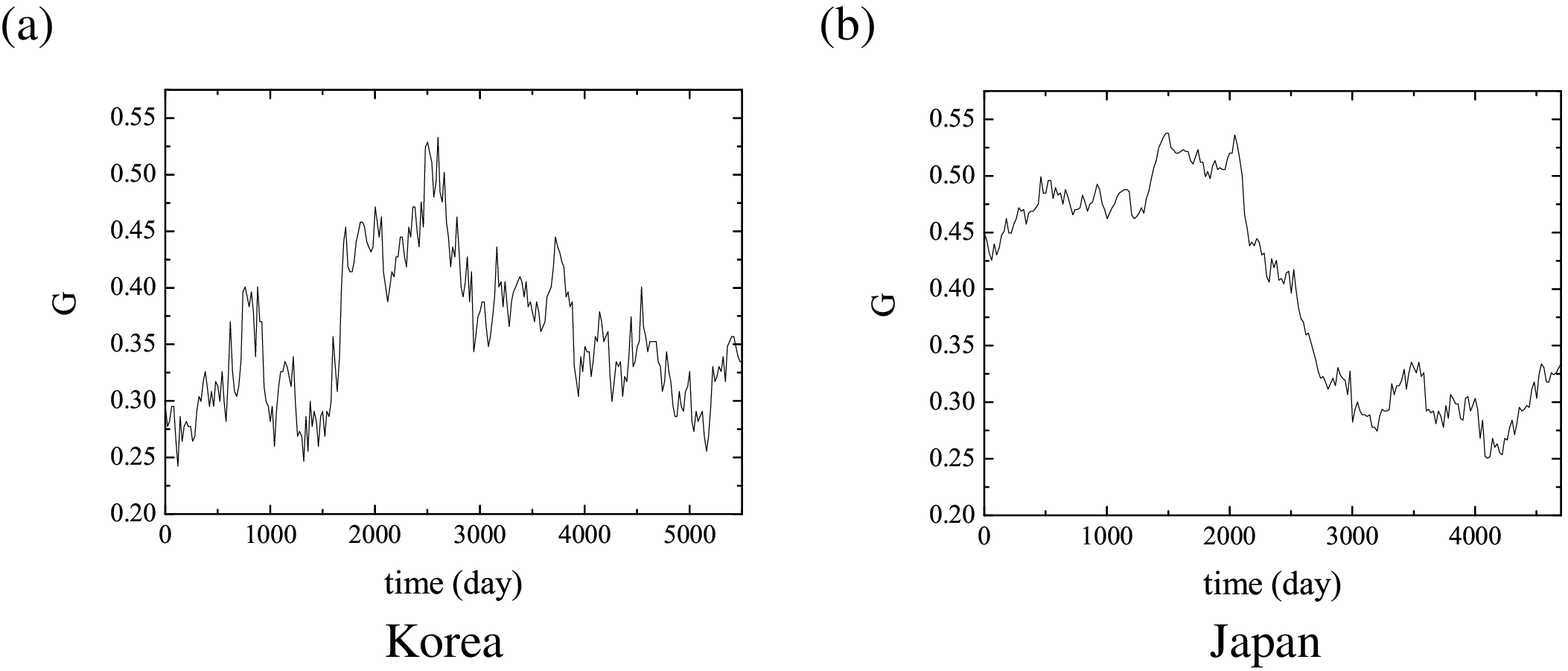}
\caption[0]{Plot of the grouping coefficient for all categories as a
function of time from 1980 to 2001.} \label{grouptime}
\end{figure}

Tables \ref{categorykorea} and \ref{categoryjapan} show 17 and 33
categories of the Korean and Japanese markets, respectively. There
are actually more categories in the market, but we selected only
categories that contain the companies used for our data analysis.
Fig. \ref{grouptime} shows \emph{global grouping coefficient} $G$
for the Korean and Japanese markets as a function of time. This
coefficient, $G$, was defined with all of the nodes, and the ratio
of connections between companies in the same category to the total
number of links. Before the mid-1980s, the Korean market was
unstable with poor liquidity, and this is one possible explanation
for the lower value in the early 1980s in Fig. \ref{grouptime}a. As
the market prospered, groups of industry categories also extensively
formed. We found that the MST of the late-1980s can be correlated to
groups of industry categories \cite{Jungjkps2006}. However,
globalization of the Korean market has progressed, which was
hastened by the 1988 Seoul Olympic Games and the 1997 Asian
financial crisis \cite{Climent2003}. In particular, globalization of
the Korean market progressed to synchronization with external
markets. In the early 2000s, the companies of the Korean market were
grouped as the MSCI Korea Index \cite{Jung2006}. In the Korean
market, the influence of foreign traders is strong, and the MSCI
Korea index is a good reference for their trading. This explains the
decreasing coefficient in Fig. \ref{grouptime}a after the mid-1980s.

We also observed the similar trend in the Japanese market (Fig.
\ref{grouptime}b). Before the mid-1980s, the coefficient $G$ of the
Japanese market showed no special movement. However, the coefficient
tends to decrease after the mid-1980s. The influence of foreign
markets is increasingly strong in the Asian market, including the
Japanese and Korean markets. In the Japanese market this tendency
was strengthened more and more by burst of the bubble in 1990's. It
is natural to form groups of industry categories, because companies
included in the same category are highly related to each other in
comparison to companies in other categories. However, the recent
globalization of Asian markets has progressed to synchronization
with foreign markets, especially the US market, and then the groups
of industry categories are breaking down.

For example, the largest company in the Korean market is Samsung
Electronic Co. (SEC), which is included in the \emph{Electrical \&
electronic equipment} category. The Korean IT industry is well
developed and there are many companies in the Electrical \&
electronic equipment category. SEC is well known throughout the
world, and many foreigners trade in SEC stock. However, other
companies in the same category are not as well known, and their
stock price is easily influenced by the economic situation in Korea.
On the other hand, SEC is synchronized to foreign markets or
factors, and is thus separated from the category. This leads to a
decrease in the grouping coefficient.

\begin{figure}
\includegraphics[angle=0,width=1.0\textwidth]{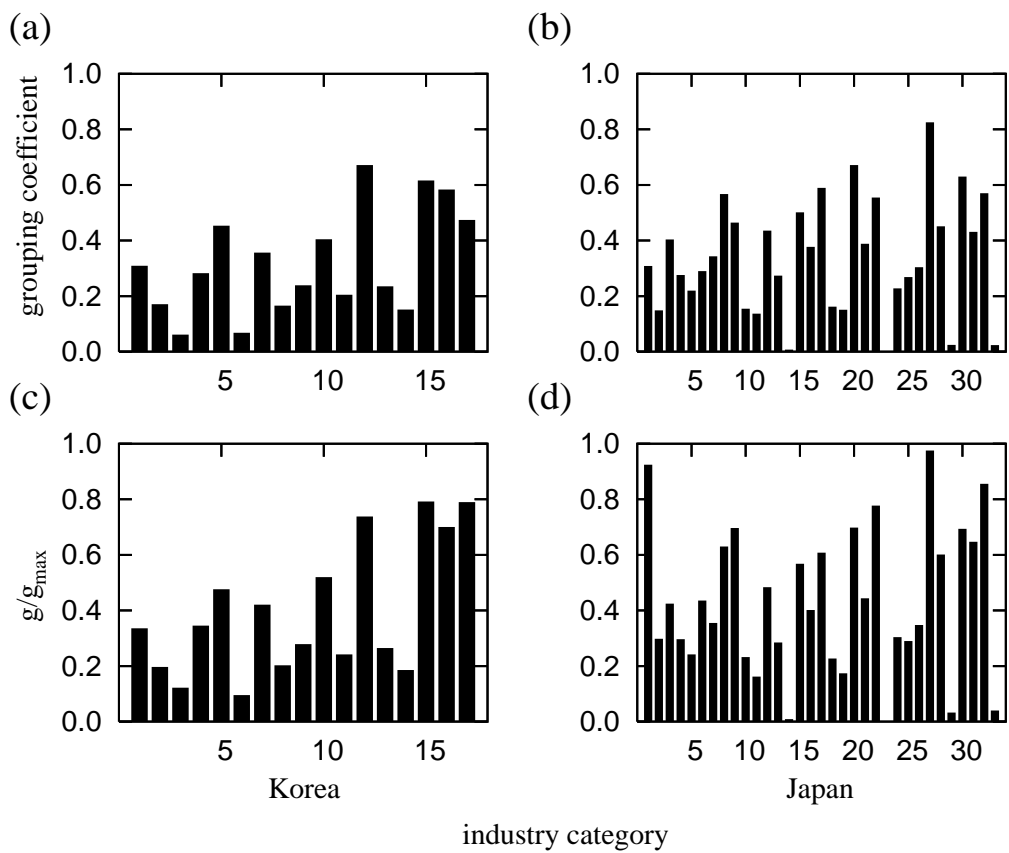}
\caption[0]{Plot of grouping coefficients: (a,b) $g$ values; (c,d)
ratio of the coefficient to the maximum value of $g$ for the Korean
and Japanese markets, respectively.} \label{group}
\end{figure}

We define the quantity \emph{grouping coefficient} to investigate
the dynamics of the grouping method in the market. The coefficient
of a given industry category $C$ is defined as:

\begin{equation}
g_{C}=\frac{\sum_{i\in C}n^C(i)}{\sum_{i\in C}n(i)}
\end{equation}

\noindent where $i_{\in C}$ are the nodes in category $C$, $n(i)$ is
the number of links connected to node $i$ and $n^{C}$ is the number
of links from the node included in category $C$. Fig. \ref{group}a
shows the grouping coefficient for each category of the Korean
market over the whole period. We can observe that categories 12, 15,
16 and 17 form a well-defined group. Category 12 is the
\emph{Construction} industry, and reflects domestic demand. We think
that categories 15 (\emph{Banks}), 16 (\emph{Insurance}), and 17
(\emph{Securities}) can be regarded as one category, the
\emph{Financial industry}. Fig. \ref{group}b shows the grouping
coefficients for the Japanese market. Categories 27, 30, 31 and 32
are well-defined categories. In addition, categories 30
(\emph{Banks}), 31 (\emph{Securities \& commodity futures}) and 32
(\emph{Insurance}) are regarded as a financial industry category. In
Japan, there are several electric distribution companies, and they
have a monopoly for a given area, e.g. Tokyo Electrical Co. and
Osaka Electrical Co. Category 27, \emph{Electric power \& gas},
consists of such companies, which are highly related to each other.
The category represents an domestic industry that is not strongly
influenced by foreign factors. Another domestic industry is category
20, (\emph{Construction}), which also forms a rather well-defined
group. Domestic industry is relatively independent of foreign
factors, and companies included in the financial category are highly
correlated to each other, regardless of foreign factors. This
explains why the coefficients for these categories are higher over
the whole period.

We now focus on the maximum grouping coefficient for each industry
category. We take the maximum value when the nodes are linked
linearly. The maximum value of the coefficient for Korean category
17 (\emph{Securities},include only four companies, is only 0.6
(=3/5) because of the properties of the MST. Fig. \ref{group}c,d
shows the ratio of the grouping coefficient to the maximum value for
each category. Fig. \ref{group} confirms that the categories
mentioned are well-defined groups. There are some errors in the
plots. For example, there are only three companies in Japanese
category 1, and its value increases when we consider the maximum
value. However, it is not meaningful because of the small sample
size.

\section{Conclusion}
\begin{figure}
\includegraphics[angle=0,width=1.0\textwidth]{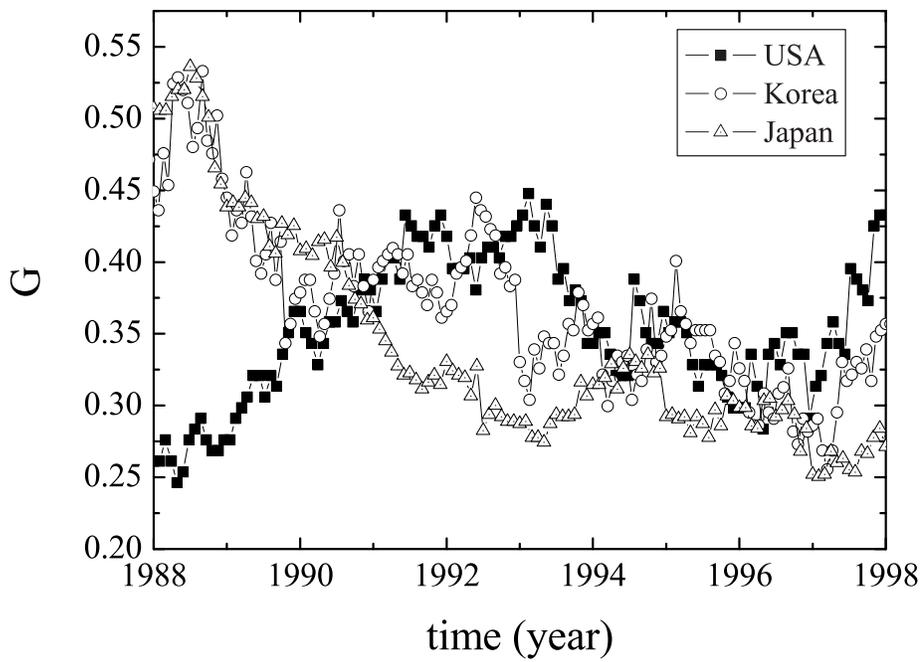}
\caption[0]{Plot of the grouping coefficient for the US, Japanese
and Korean markets.} \label{grouptimefinal}
\end{figure}

We investigated the Japanese and Korean stock market networks using
the daily closure stock price. Our analysis shows that the grouping
coefficients of the two markets decreased with elapsing time and the
number of groups according to industry categories decreased. During
the same period, the coefficient of the US market did not decrease
(Fig. \ref{grouptimefinal}). Currently, most world markets,
especially the Japanese and Korean markets, synchronize to the US
market, and they are sensitive to foreign factors. The grouping
coefficient represents a good parameter for measuring this
phenomenon. Our future work will involve analysis of other markets,
which should confirm the usefulness of the grouping coefficient.
Also, we confirmed that the MST is a good analysis tool for
investigating the stock market.

\begin{ack}
We wish to thank J.-S. Yang for active discussion. We also thank W.
Lee, G. Oh, W.C. Jun, and S. Kim for useful support.
\end{ack}

\newpage
\clearpage

\newpage
\clearpage

\newpage
\clearpage

\newpage
\clearpage

\newpage
\clearpage
\begin{table}
\caption{Industry categories of the Korea Stock Exchange in our data
set} \label{categorykorea}
\begin{tabular}{|c|c|}
\hline
\#&Industry category\\
\hline
1&Food \& beverages\\
2&Textiles\\
3&Apparel\\
4&Paper \& wood\\
5&Chemicals \& medical supplies\\
6&Rubber\\
7&Non-metallic minerals\\
8&Iron \& metals\\
9&Manufacturing \& machinery\\
10&Electrical \& electronic equipment\\
11&Transport equipment\\
12&Construction\\
13&Distribution\\
14&Transport \& storage\\
15&Banks\\
16&Insurance\\
17&Securities\\
\hline
\end{tabular}
\end{table}

\newpage
\clearpage
\begin{table}
\caption{Industry categories of the Tokyo Stock Exchange in our data
set} \label{categoryjapan}
\begin{tabular}{|c|c||c|c|}
\hline
\#&Industry category&\#&Industry category\\
\hline
1&Fishery, agriculture \& forestry&18&Information \& communication\\
2&Mining&19&Other products\\
3&Foods&20&Construction\\
4&Textiles \& apparel&21&Land transportation\\
5&Glass \& ceramics products&22&Marine transportation\\
6&Pulp \& paper&23&Air transportation\\
7&Chemicals&24&Warehousing \& harbor transportation services\\
8&Pharmaceuticals&25&Wholesale trade\\
9&Oil \& coal products&26&Retail trade\\
10&Rubber products&27&Electric power \& gas\\
11&Precision instruments&28&Real estate\\
12&Iron \& steel&29&Services\\
13&Machinery&30&Banks\\
14&Metal products&31&Securities \& commodity futures\\
15&Non-ferrous metals&32&Insurance\\
16&Transportation equipment&33&Other Financing business\\
17&Electrical appliances&&\\
\hline
\end{tabular}
\end{table}

\end{document}